\documentclass[abbrv,aps,prb,twocolumn,showpacs,preprintnumbers,amsmath,amssymb,
groupaddress,nofootinbib]{revtex4}

\usepackage{amsmath,amssymb,amsfonts}
\usepackage{bm,graphicx,color}

\newcommand{\laco}{$\mathrm{LaCoO_3 }$}

\begin{document}
\title{Spin state transition and covalent bonding in LaCoO$_3$}
\author{V.~K\v{r}\'apek}
\email{krapek@fzu.cz}
\author{P.~Nov\'ak}
\author{J.~Kune\v{s}}
\affiliation{Institute of Physics, Academy of Sciences of the Czech Republic,
Cukrovarnick\'a 10,
Praha 6, 162 53, Czech Republic}
\author{D.~Novoselov}
\author{Dm.~M.~Korotin}
\author{V.~I.~Anisimov}
\altaffiliation{Also at Ural Federal University, 620002 Yekaterinburg, Russia.}
\affiliation{Institute of Metal Physics, Russian Academy of Sciences, S. Kovalevskaya 18, 620990 Yekaterinburg, Russia}
\date{\today}

\begin{abstract}
We use the dynamical mean-field theory to study a $p$-$d$ Hubbard Hamiltonian
for LaCoO$_3$ derived from {\it ab initio} calculations in local density
approximation (LDA+DMFT scheme). We address the origin of local 
moments observed above 100~K and discuss their attribution to a particular
atomic multiplet in the presence of covalent Co-O bonding.
We show that in solids such attribution, based on the single ion picture, is in general 
not possible.
We explain when and how the single ion picture can be generalized to provide 
a useful approximation in solids. Our results demonstrate that the apparent 
magnitude of the local
moment is not necessarily indicative of the underlying atomic multiplet.
We conclude that the local moment behavior in LaCoO$_3$ arises from the high-spin
state of Co and explain the precise meaning of this statement.

\end{abstract}
\maketitle

\section{Introduction}
\laco\ and related compounds have been much studied for
half a century.\cite{Heikes19641600,naiman:1044,raccah}
Their strongly temperature dependent magnetic and
transport properties have been eluding complete theoretical description
so far.\cite{korotin,PhysRevB.67.172401,PhysRevB.71.054420,PhysRevB.77.045130,PhysRevB.77.045123,Hsu,eder}
An apparent band insulator below 50~K, \laco\ exhibits a local moment magnetic
response above 100~K while the charge gap continuously disappears
between 450 and 600~K.\cite{PhysRevB.6.1021,saitoh,ISI:000176767100012} This behavior points 
to an important role played by the electronic correlations as common
among transition metal oxides. The correlated nature of \laco\ reveals 
itself also in the formation of atomic-scale inhomogeneities
with large magnetic moments upon moderate hole doping.\cite{PhysRevLett.101.247603,PhysRevB.83.134430}
The picture of thermal evolution of \laco\ as
an entropy driven crossover from a non-magnetic to a magnetic state of Co$^{3+}$ 
ion has been generally accepted.
However, the actual why's and how's are far from settled.
The main open questions include the following. (i) Which atomic multiplet is responsible
for the formation of local moments? 
(ii) Does the lattice thermal expansion actively contribute to the spin-state
transition or is it merely a slave to the changes of the electronic structure?
(iii) Why do the crossovers to local moment paramagnet and to bad metal
happen at different temperatures?
In this work we use the combination of the density functional band structures
with the dynamical mean-field theory, known as LDA+DMFT,~\cite{PhysRevLett.62.324,PhysRevB.45.6479,RevModPhys.68.13,0953-8984-9-35-010,PhysRevB.57.6884,0953-8984-20-6-064202}
to address the former two questions. We discuss in detail the attribution
of local moment behavior to a particular atomic multiplet in systems with
covalent bonds, a question of general importance in oxide physics.

The magnetic susceptibility of \laco\ is usually analyzed
in terms of the lowest multiplets of an isolated Co$^{3+}$ ion in octahedral        
crystal field:\cite{tanabe-sugano}
the low-spin (LS) $^1A_1$ ($t^6e^0$), the intermediate-spin (IS) $^3T_1$
($t^5e^1$), or the high-spin (HS) $^5T_2$ ($t^4e^2$) states.
The energy differences between the multiplets  
are controlled by the crystal-field (CF)
splitting and the intra-atomic exchange $J$.
While the LS singlet ground state is undisputed (at least for low temperature
crystal structures)
the nature of the first excited state is still a subject of debate.
Goodenough~\cite{Goodenough1958287} attributed the appearance of 
local moment to population of the HS state.
Heikes {\it et al.},~\cite{Heikes19641600} on the other hand, proposed
the IS scenario, which became popular~\cite{radaelli,asai,zobel}
after Korotin {\it et al.}~\cite{korotin} obtained
an IS ground state for expanded lattice with LDA+U
calculation, contrary to a simple ligand-field theory.
More recent experiments make a strong case for the HS scenario.
The electron spin resonance shows a triplet excited state
with a large $g$ factor of 3.35\,--\,3.55 which is consistently explained
in the HS scenario invoking the effect of spin-orbit coupling.\cite{PhysRevB.66.094404,
PhysRevB.67.172401} The x-ray absorption spectra (XAS) and
magnetic circular dichroism at the $L_{2,3}$ edge of Co \cite{haverkort}
also select the HS scenario. 
However, several authors \cite{kyomen1,kyomen2,haverkort,eder} pointed out 
that in order to interpret the magnetic susceptibility, specific heat,
or XAS data in the HS scenario a rather strong temperature dependence
of the crystal field has to be assumed. In other words, the experimentally
deduced increase of the HS population is significantly slower than anticipated with a fixed
crystal field. The implication is that the apparent crystal field
grows with temperature. This is somewhat unexpected since the expansion
of the Co-O bond which accompanies the spin crossover
should reduce the crystal field. A possible explanation of this
puzzling feature is provided by an interatomic repulsion between the HS states, which
is equivalent to attraction between HS and LS states.~\cite{kyomen1,kyomen2,eder}
Breathing lattice distortion proposed by Raccah and Goodenough,~\cite{raccah}
studied in detail by Bari and Sivardi\`ere,~\cite{bari72} provides
a mechanism. Kn\'{i}\v{z}ek {\it et al.} \cite{knizek1} used LDA+U calculations to argue for 
HS-LS attraction. Recently, a purely electronic mechanism of HS-LS attraction
was observed by two of us \cite{krapek} in a two band Hubbard model
and by Zhang {\it et al.}~\cite{zhang} in \laco\ specific calculation.

It is well known that the effective crystal field in transition
metal oxides is largely due to hybridization with ligands, 
i.e.,~generated by hopping of predominantly $e_g$ electrons between
the metal and oxygen sites. This effect is particularly pronounced in \laco\
and leads to charge fluctuations on the Co site. 
Therefore, it seems natural to question descriptions based on an isolated ion. 
How does one define and distinguish HS and IS states when sizable charge (valence) 
fluctuations are present?
Or, more precisely, is it possible to express the $T$-dependent susceptibility as a
sum of contributions from different atomic states? 
We will show in Sec.~\ref{method_susceptibility} that in a solid the answer is negative in general.
However, in the insulating phase of \laco\ the notion of LS, IS, and HS states can be preserved
when these are generalized to include the hybridization induced charge fluctuations. 
The $T$-{\it dependent} paramagnetic moment
can be to a good accuracy approximated by a $T$-{\it
independent} HS moment multiplied by
a $T$-{\it dependent} weight. Importantly, the magnitude of this
effective moment differs from the free ion value.

The role of the lattice thermal expansion poses a chicken-and-egg question
about the relationship between the spin-state transition
and the anomalous lattice expansion.~\cite{PhysRevB.50.3025,zobel,PhysRevB.78.134402} While the transition to
the magnetic state, both IS and HS,
weakens the Co-O bond and thus leads to its expansion, stretching the Co-O
bond reduces the effective
CF splitting and thus favors a magnetic state. Therefore, there is a
positive feedback between these two effects.
To include the lattice response directly into our calculations is not
computationally feasible at the moment.
Therefore, we have performed calculations for several lattices
corresponding to experimental
crystal structures at different temperatures. By comparing the
$T$ dependence of the spin susceptibilities
on these lattices we find that experimentally observed variation of Co-O
bond lengths has a pronounced effect
on the electronic properties.

Inclusion of thermal effects
in ``first principles" density functional approaches is notoriously difficult.
Therefore, the existing studies are either limited to the $T=0$ LS phase~\cite{abbate_lda}
or assume that temperature enters only through the lattice thermal
expansion.~\cite{korotin,knizek1} The only serious attempt to explicitly include temperature 
in such a calculation was made by Eder~\cite{eder} using the variational cluster
approximation (VCA). The VCA and DMFT methods share many formal similarities,
but involve different approximations.
VCA is, in principle, an exact method for calculation
of one-particle properties, but, as pointed out in Ref.~\onlinecite{eder},
the relevance of multiplets populations obtained
from a reference CoO$_6$ cluster is a conjecture,
which calls for verification with other methods.
DMFT, on the other hand, treats one-particle
and multiparticle correlations on the same footing,
but becomes exact only in the limit of vanishing
nonlocal correlations (infinite dimension).
Recently, DMFT was applied to \laco\ to study
the effect of varying interaction strength and pressure.~\cite{zhang}

\section{Methods}
\label{method}
\subsection{Model}
Multiband Hubbard Hamiltonian with the two-particle interaction within 
the Co $3d$ shell is used to describe \laco.
The one-particle part of the Hamiltonian, which spans the 
Co $3d$ and O $2p$ orbitals, has been constructed from the
local density approximation (LDA) to the density functional theory.
The non-spin-polarized band structure obtained with 
\textsc{wien}2k~\cite{wien2k} was transformed into the
Wannier basis representation with \textsc{wien2wannier}~\cite{w2w} and
\textsc{wannier}90~\cite{wannier90} codes. The Hamiltonian in this representation reads
\begin{equation}
\begin{split}
H=\sum_{\bm{k}\sigma}\left( 
h_{\bm{k},\alpha\beta}^{dd} d_{\bm{k}\alpha\sigma}^\dagger d_{\bm{k}\beta\sigma}
+
h_{\bm{k},\gamma\delta}^{pp} p_{\bm{k}\gamma\sigma}^\dagger p_{\bm{k}\delta\sigma}
\right .
\\
+\left .
h_{\bm{k},\alpha\gamma}^{dp} d_{\bm{k}\alpha\sigma}^\dagger p_{\bm{k}\gamma\sigma} 
+
h_{\bm{k},\gamma\alpha}^{pd} p_{\bm{k}\gamma\sigma}^\dagger d_{\bm{k}\alpha\sigma}
\right)
+
\sum_{i,\sigma,\sigma^\prime} U_{\alpha\beta}^{\sigma\sigma^\prime}
n^d_{i\alpha\sigma} n^d_{i\beta\sigma^\prime}.
\end{split}
\label{Hamiltonian}
\end{equation}
Here $d_{\bm{k}\alpha\sigma}$ and $p_{\bm{k}\gamma\sigma}$ are
Fourier transforms of the annihilation operators
$d_{i\alpha\sigma}$, $p_{i\gamma\sigma}$ which destroy
the $d$ or $p$ electron with the orbital index $\alpha$ or $\gamma$
and the spin index $\sigma$ in the $i$th unit cell,
$n^d_{i\alpha\sigma}$ are the corresponding occupation
number operators, and $h_{\bm{k},\alpha\beta}^{ab}$ are the
corresponding matrix elements of the one-particle LDA Hamiltonian.

The $U_{\alpha\beta}^{\sigma\sigma^\prime}$ are the density-density 
matrix elements in the $e_g$-$t_{2g}$ basis of the full Coulomb interaction~\cite{PavariniChapter6}
parametrized with $U$ (Slater parameter $F_0$) and Hund's exchange $J$
[connected with the Slater parameters $F_2$, $F_4$ as $J=(F_2+F_4)/14$,
$F_4/F_2=0.625$].
The screened values of $U=6.0$~eV and $J=0.8$~eV
have been obtained using the constrained density functional theory (cDFT), 
described in detail in Ref.~\onlinecite{springerlink:10.1140/epjb/e2008-00326-3}. 
To the DFT potential an orbitally dependent term was added, which shifts 
by a small amount the energy of
selected Wannier functions (WFs). Due to this shift, occupation of the WFs
in question is changed and the Coulomb
interaction parameters are then determined as a derivative of the site energy with 
respect to the occupation.

In the following calculations are performed with cDFT values
of $U$ and $J$.
Calculations, where $U$ and $J$ were varied, are also reported to assess 
the stability of the results.

The $h^{dd}$ diagonal elements were modified to account for the static
part of the interaction, double-counting correction,
\begin{equation}
\label{eq_edc}
h_{\bm{k},\alpha\beta}^{dd}=h_{\bm{k},\alpha\beta}^{0,dd}-
(N_\mathrm{orb}-1)\bar{U}\bar{n}\delta_{\alpha\delta},
\end{equation}
where $\bar{n}$ is the average self-consistent occupancy per Co:$d$ orbital,
$\bar{U}$ is the orbital averaged interaction energy,
and $N_\mathrm{orb}$ is the total number of interacting orbitals
on a single site (10 in our case).\cite{fe2o3} This is equivalent to subtracting the 
orbitally averaged Hartree potential felt by the $d$ electrons.

\subsection{DMFT calculations and one-particle spectra}
The one-particle Green's function of the Hamiltonian (\ref{Hamiltonian}) 
is found by iteratively solving the DMFT equations on the Matsubara contour.
The auxiliary impurity problem is solved
by the continuous time quantum Monte Carlo (CT-QMC) method in
the hybridization expansion formulation~\cite{Werner}
using the implementation based on free-access package ALPS.~\cite{alps,ALPS2}
The Wang-Landau reweighting~\cite{PhysRevLett.86.2050,PhysRevE.64.056101}
was employed in order to ensure the ergodicity of the simulations for some parameter values,
in particular at low temperatures and close to the spin state transition.

Once the calculation was converged we have evaluated the one-particle spectra
in real frequency and analyzed the impurity dynamics and spin susceptibility.
For the former analytic continuation is necessary. To this end we have employed
the maximum entropy method in two modes: (i) continuation of the local Green's function
from the imaginary time $\tau$ to real frequency $\omega$ and (ii) continuation
of the local self-energy from the Matsubara frequency $i\omega_n$ to $\omega$.
For the latter we have used the statistical error estimates following Ref.~\onlinecite{PhysRevB.80.045101}.
The former was used to cross-check the results of (ii) and the spectra are not shown in the paper.
Analytic continuation of self-energy has several attractive features, such as
being exact in the noninteracting limit, providing a direct access
to the $k$-resolved and ligand spectra, and smearing out the lifetimes (imaginary
part of the self-energy) but not the quasiparticle dispersions.

\subsection{Susceptibility and local moments}
\label{method_susceptibility}
To analyze the on-site dynamics we have studied two additional quantities.
First, the state weights \cite{PhysRevLett.99.126405} which are the diagonal terms of the
site-reduced density matrix, i.e., expectation values of the projection operators
on the atomic (many-body) states $\hat{P}_\mu=|\mu\rangle\langle \mu|$. 
In the present case of the density-density interaction the site-reduced density matrix
is diagonal in the occupation number basis. Therefore, the knowledge of 
the state weights allows us to evaluate the expectation value of any local operator, e.g.,
the instantaneous local moment $\langle \hat{m}_z^2
\rangle=\sum_{\mu}m_z^2(\mu)\langle \hat{P}_{\mu} \rangle$ with
$m_z(\mu)=\langle\mu|\hat{m}_z|\mu\rangle$. 

Second, we define the imaginary time 
state-state correlation matrix $C_{\mu\nu}(\tau)$ and its time average $\Pi_{\mu\nu}$,
\begin{equation}
\begin{split}
C_{\mu\nu}(\tau)=\langle \hat{P}_\mu(\tau) \hat{P}_\nu(0) \rangle, \\
\Pi_{\mu\nu}=T\int_0^{\beta}d\tau C_{\mu\nu}(\tau),
\end{split}
\label{CPi}
\end{equation}
where $\beta=1/T$ is the inverse temperature.
The correlation matrix allows us to analyze the local response functions (via
fluctuation-dissipation theorem) and decompose them into different contributions.
In particular, we will be interested in the local spin susceptibility $\chi$,
which in the paramagnetic state can be expressed as 
\begin{equation}
\begin{split}
\label{eq_chi}
\chi&=\int_0^{\beta}d\tau\langle m_z(\tau)m_z(0)\rangle \\
 &=\frac{1}{T}\sum_{\mu,\nu}m_z(\mu)m_z(\nu)\Pi_{\mu\nu}.
\end{split}
\end{equation}
The second expression shows that in general the local magnetic response cannot be 
decomposed into contributions of atomic states, but pairs of states must be considered.
Decomposition into individual states contributions is possible only if $C_{\mu\nu}(\tau)$ can be made diagonal, e.g., in an isolated atom.

Discussing briefly the physical meaning of these quantities we start by pointing out
that in the course of time a given atom visits various quantum-mechanical states
as a result of statistical (thermal) fluctuations and quantum-mechanical (causal)
evolution. The weight of a given state is a relative measure of the time spent
by the atom in this state, which does not distinguish between thermal fluctuations
and causal evolution. The state-state correlations distinguish to some extent
between these two effects as only states connected by causal evolution can have
a nonzero cross term. The state weights can be obtained as row (or column)
sums over $\Pi_{\mu\nu}$.

\section{Results and discussion}
\label{results}

\subsection{Non-interacting band structure}
We have considered the experimental distorted
perovskite structure with $R\bar{3}c$ space group containing
two formula units per unit cell. The structural parameters were taken
from the x-ray measurements of Ref.~\onlinecite{radaelli}. To assess the effect of
lattice thermal expansion the calculations were repeated for the experimental
structural parameters obtained at three different temperatures (denoted as
$\tau_\mathrm{lattice}$ in the following):
5, 450, and 750~K.

\begin{figure}[ht]
\includegraphics[angle=270,width=\columnwidth,clip]{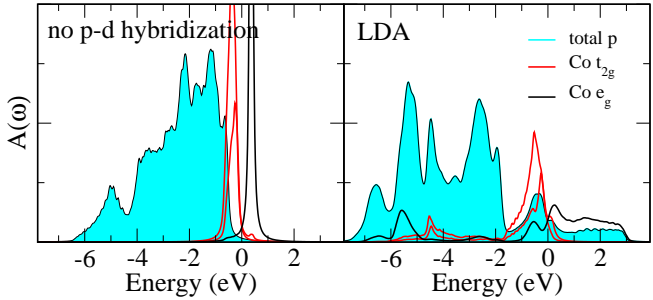}
\caption{
(Color online) Calculated orbitally resolved spectral function
[O:$p$ (shaded cyan); Co:$t_\mathit{2g}$ and $e_g$ 
(red and black line)].
Left panel: the hybridization term $h_{\bm{k},\alpha\gamma}^{dp}$ of the Hamiltonian (\ref{Hamiltonian})
was set to zero. Right panel: the LDA Hamiltonian in the Wannier basis.
}
\label{fig_spec_hyb}
\end{figure}

The octahedral crystal field splits Co:$d$ states into six $t_{2g}$ states
at lower energy and four $e_g$ states at higher energy.
The resulting orbitally resolved spectral density is shown
in Fig.~\ref{fig_spec_hyb}. The splitting is strongly
contributed by the O:$p$--Co:$d$ hybridization, as is clear from the comparison
in Fig.~\ref{fig_spec_hyb}, where the right and left panels show the spectral
function with and without the $p$-$d$ hybridization included.
The on-site contribution to the crystal field splitting $\Delta$ (left panel of Fig.~\ref{fig_spec_hyb})
is close to the value 0.7~eV extracted from the XAS measurements.~\cite{haverkort}
The $p$-$d$ hybridization increases
the distance between the centers of $t_{2g}$ and $e_g$ bands considerably.
The band broadening, more pronounced for the $e_g$ band, is another
consequence.
Matching O:$p$ and Co:$d$ features reflect formation of bonding and
antibonding states.
Stronger hybridization of the $e_g$ orbitals compared to $t_{2g}$ ones
results from their larger spatial overlap with O:$p$ orbitals. The $t_{2g}$
band is further split to the $e_g^\pi$ doublet and the $a_{1g}$ singlet due
to a distortion from the octahedral symmetry. This splitting does not
play an important role in our study, though.
As in previous calculations a metallic ground state 
is incorrectly predicted by LDA.~\cite{Hsu}

\subsection{Thermal effects and lattice expansion}
LDA+DMFT calculations were performed for several temperatures between 290 and 2320~K
($\beta$ from 40 to 5~eV$^{-1}$).
If not stated otherwise, the results are shown for $U=6.0$\,eV and $J=0.8$\,eV
(obtained by cDFT calculations).
\begin{figure}[ht]
\includegraphics[width=\columnwidth,clip]{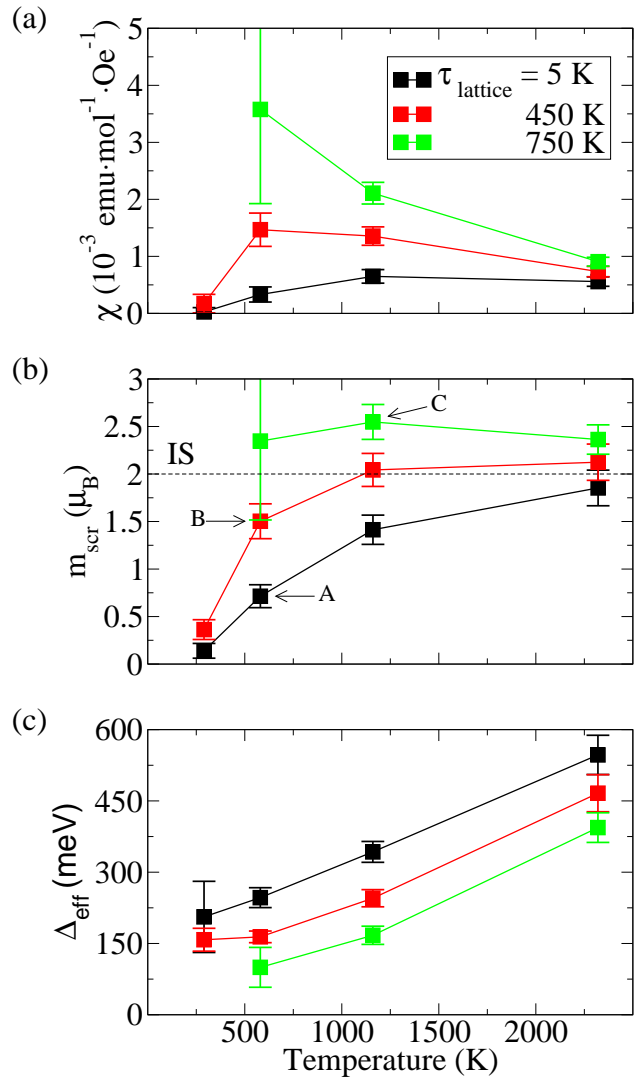}
\caption{
(Color online)
(a) Local spin susceptibility as a function
of the temperature for different lattices ($\tau_\mathrm{lattice}$).
(b) The corresponding screened spin moment. 
The dashed line indicates the value for the IS in the ionic limit.
The letters A, B, C denote the solutions discussed in the text.
(c) The apparent crystal field $\Delta_{\text{eff}}$ obtained
from Eq.~(\ref{eq_cfeff}).
}
\label{fig_chi_mtm}
\end{figure}

{\it Local susceptibility.}
The local spin susceptibility $\chi$, calculated
from Eq.~(\ref{eq_chi}), is shown in
Fig.~\ref{fig_chi_mtm}(a) as a function of $T$.
For $\tau_\mathrm{lattice}=750$~K the huge error bar 
at $T=580$~K is due to a long autocorrelation time despite the
Wang-Landau~\cite{PhysRevLett.86.2050,PhysRevE.64.056101} sampling.
For all lattice parameters we observe an emergence of Curie-like susceptibility at high temperatures.
The corresponding average spin moments $m_\mathrm{scr}=\sqrt{T\chi}$
are shown in Fig.~\ref{fig_chi_mtm}(b). As we calculate directly the local
susceptibility (the local response to a field applied to a single site of the infinite
crystal) the intersite exchange does not enter the definition of the local moment.
Figure~\ref{fig_chi_mtm}(b) suggests a gradual thermal population of a magnetic
state. The lattice expansion clearly favors the magnetic state and the experimentally
observed magnitude of the Co-O bond-length expansion has a sizable impact on our results.

Next, we discuss the temperature dependences $\chi(T)$, obtained from Eq.~(\ref{eq_chi}), for a fixed lattice.
To get in touch with experimental observations we adopt the single ion
expression commonly used in analysis of experimental data,
\begin{equation}
\label{eq_cfeff}
\chi(T)=\frac{\mu^2}{T}
\frac{\nu}{\nu+\exp[\Delta_{\text{eff}}(T)/T]},
\end{equation}
where $\mu$ and $\nu$ are the magnitude of the local moment and the multiplicity of 
the excited magnetic state and
$\Delta_{\text{eff}}$ is the excitation energy with respect to the LS ground state.
In Fig.~\ref{fig_chi_mtm}(c) we show $\Delta_{\text{eff}}$ obtained
from Eq.~(\ref{eq_cfeff}) using $\nu=6$ and $\mu=3.5$, which correspond to 
an Ising HS state, a choice explained later in the text. Like in the experiments \cite{kyomen1,kyomen2,haverkort}
and the VCA theory,~\cite{eder} our $\Delta_{\text{eff}}$ for a fixed lattice increases with temperature
for the reason discussed below.
In particular, the increase of $\Delta_\mathrm{eff}$
by a factor of 3--5 over the spin-state crossover was deduced from
XAS~\cite{haverkort} and measurements of the magnetic susceptibility and heat
capacity.~\cite{kyomen2}

It is quite clear that our results do not provide an accurate quantitative description
of \laco\ as the spin crossover takes place at a too high temperature. This is not surprising
since the present theory is unlikely to achieve the necessary $\sim$10~meV accuracy without 
fine-tuning the material
parameters by hand.
The approximations of the model (such as the restriction to
the density-density Coulomb interaction and neglect of the long-range
or $p$-$d$ interactions) limits the accuracy further.
However, two important trends are revealed. First, the lattice response (expansion
of O$_6$ octahedra around moment-carrying sites) acts as a positive feedback for generation
of local moments. Second, this is countered by a purely electronic effect,
making addition of a local moment the harder the more local moments are
already in the system; this is reflected in the increase of
$\Delta_{\text{eff}}$ with the temperature [Fig.~\ref{fig_chi_mtm}(c)].
This is another way of saying that there is a repulsive interaction between the magnetic
sites in the LS background.
We point out that in our calculation all O$_6$ octahedra are the same, which excludes
a possible contribution of the breathing distortion.~\cite{raccah} Indeed, Ky\^omen {\it
et al.}~\cite{kyomen1} substituting Co with Al, Ga, and Rh came to the conclusion that electronegativity 
rather than ionic radius of the neighbors is the parameter which controls $\Delta_{\text{eff}}$.
We suggest the following picture based on the observation that
a strong Co:$e_g$-O bond favors LS state, and that in the Co-O-Co trimer the Co-O bonds share the
central $p_{\sigma}$ orbital. Due to this sharing the energy gain per Co-O bond in the trimer
is less than the energy gain for an isolated Co-O bond. Therefore, breaking (weakening) one bond 
in the trimer makes the other bond stronger. Introducing a local moment on one Co site provides
this bond breaking and strengthening the other bond favors the LS state.

\begin{figure}[ht]
\includegraphics[width=\columnwidth,clip]{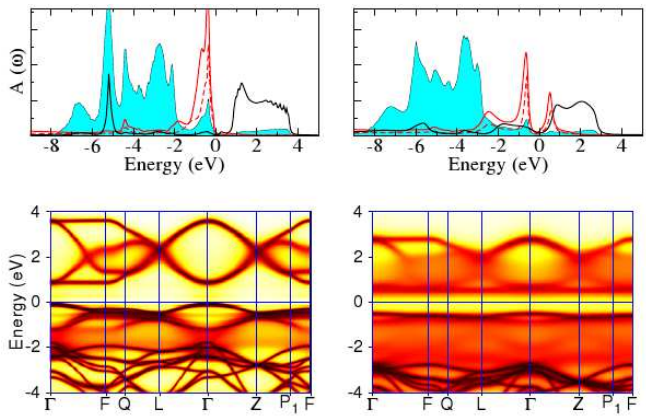}
\caption{
(Color online)
(Upper panels) Orbital-resolved spectral functions 
(states/eV/formula unit). $t_\mathit{2g}$ ($e^{\pi}_g$): solid red line;
$t_\mathit{2g}$ ($a_\mathit{1g}$): dashed red line; $e_g$: black line;
O:$p$: cyan shaded area. The O:$p$ spectral function is downscaled
by the factor of 2 to fit in the graph.
(Lower panels) $k$-resolved spectral function $A_\mathbf{k}(\omega)$ along
the high symmetry directions.
Left panels show the interacting
nonmagnetic (low-spin) solution [denoted as A in Fig.~\ref{fig_chi_mtm}(b);
$T=580$~K, $\tau_\mathrm{lattice}=5$~K] and the right panels
display the paramagnetic solution (with a large content of high-spin atomic states)
[denoted as C in Fig.~\ref{fig_chi_mtm}(b); $T=1160$~K; $\tau_\mathrm{lattice}=750$~K]
.
}
\label{fig_spectral_k}
\end{figure}

{\it Spectral functions.}
In Fig.~\ref{fig_spectral_k} we compare the one-particle spectra of the low-$T$ nonmagnetic state (left panels)
and the high-$T$ paramagnetic state (right). The orbital resolved spectra are displayed in the upper panels
and the $k$-resolved spectra along the high symmetry directions are
shown in the lower panels.
The nonmagnetic spectrum resembles the LDA solution (Fig.~\ref{fig_spec_hyb}), the main difference being a uniform (Hartree) shift
of the $e_g$ band. There is very little dynamical renormalization since the LS state is
an approximate eigenstate of both the kinetic term (dominated by the crystal field) and the interaction term
taken separately. The correlated nature of \laco\ is revealed at elevated temperature.
The thermal population of the excited atomic multiplets leads to a formation of local moments, which
are incompatible with dispersive bands. As a result incoherent features appear in the spectra. The nature
of the charge gap changes from a semiconductor like gap between coherent valence $t_{2g}$ and conduction
$e_g$ bands (left panel of Fig.~\ref{fig_spectral_k}) to a $t_{2g}$-$t_{2g}$ gap (right panel of Fig.~\ref{fig_spectral_k}).
The bottom of the conduction band is now defined by an incoherent
$t_{2g}$ excitation, the tail of which gradually fills the gap with the increasing temperature.
The top of the valence manifold is formed by a renormalized dispersive $t_{2g}$ band. This
spectrum is consistent with the positive Seebeck coefficient \cite{senaris,jirak} indicating
holes to dominate the electronic transport. 
Like the VCA results,~\cite{eder} the photoemission part of the spectra exhibits a transfer
of spectral weight from the low-energy peak ($\sim$1~eV) to higher energies ($\sim$3~eV)
observed experimentally.~\cite{abbate}

\begin{figure}[ht]
\includegraphics[angle=270,width=\columnwidth,clip]{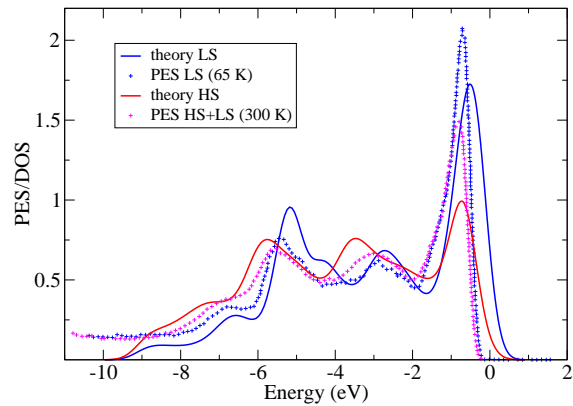}
\caption{
(Color online)
Comparison of the calculated density of states (lines) with the PES measurements (symbols). The calculated density denoted as LS is taken for $T=580$~K, $\tau_\mathrm{lattice}=5$~K [A in Fig.~\ref{fig_chi_mtm}(b)] and that denoted as HS
for $T=1160$~K, $\tau_\mathrm{lattice}=750$~K [C in Fig.~\ref{fig_chi_mtm}(b)].
The measurements were taken at 65 K (denoted as LS) and 300 K (denoted as HS+LS,
as the temperature is not high enough for the full spin-state crossover).
}
\label{fig_XASPES}
\end{figure}

In Fig.~\ref{fig_XASPES} we compare the calculated spectral functions to the photoemission spectra (corrected for surface
effect) of Ref.~\onlinecite{Koethe} (Fig.~2.8).
We have tuned the relative weights of the O:$p$ and Co:$d$ (1:5) spectra to
mimic the
effect of different absorption cross sections and added a Gaussian
broadening of 0.2 eV
to account for the experimental resolution. We find a good match
of the major spectral features. We also observe consistent trends
in both the O:$d$ and Co:$d$ parts of the spectra.  The more pronounced
difference
of the two theoretical spectra reflects most likely a higher degree of
LS to HS crossover.


\subsection{Spin state analysis}
\begin{figure}[ht]
\includegraphics[width=\columnwidth,clip]{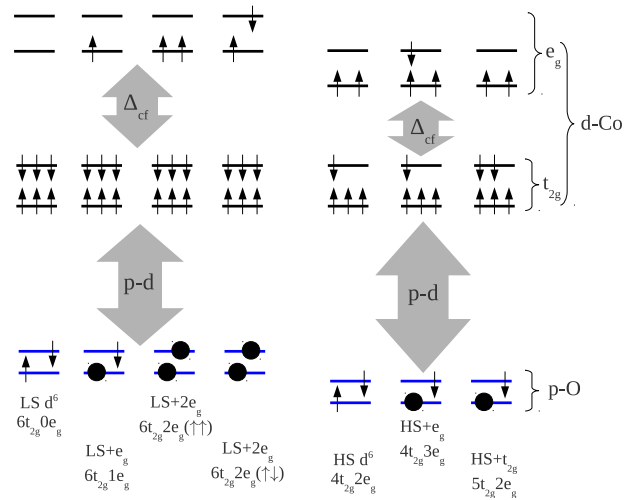}
\caption{
(Color online)
Atomic states at LS and HS configurations.
The blue lines depict the oxygen $p$ orbitals and the black 
circles denote the hole in O:$p$ shell. The effects
of the crystal field and Co:$d$-O:$p$ hybridization
are schematically depicted.
In total four atomic states belong to LS block
(see the left part of figure): (LS) all $t_\mathit{2g}$ orbitals occupied, all $e_g$ orbitals
empty, (LS+$e_g$) in addition to LS single $e_g$ orbital occupied,
[LS+2$e_g$($\uparrow\downarrow$)] in addition to LS two $e_g$ orbitals with
the opposite spin orientation occupied, and
[LS+2$e_g$($\uparrow\uparrow$)] in addition to LS two $e_g$ orbitals with
the parallel spin orientation occupied.
All these states are accessible from LS via the $e_g$ hybridization.
For brevity we do not distinguish
LS+2$e_g$($\uparrow\downarrow$) from LS+2$e_g$($\uparrow\uparrow$)
from now on.
In total three atomic states belong to HS block
(see the right part of figure): (HS) all majority spin orbitals and one minority spin 
$t_\mathit{2g}$ occupied, (HS+$e_g$) in addition to HS another minority spin
$e_g$ orbital occupied, and (HS+$t_\mathit{2g}$) in addition to HS another
minority spin $t_\mathit{2g}$ orbital occupied.
All these states are accessible from HS via the $e_g$ or $t_\mathit{2g}$
hybridization. 
}
\label{fig_splitting}
\end{figure}

{\it Local state statistics.}
Next, we address the local dynamics at Co sites and
whether it is meaningful to describe it in terms of the 
atomic states (such as LS, IS, and HS). The hybridization
expansion CT-QMC solver is well suited to this task as it
provides the site-reduced statistical operator (density matrix),
referred to as state statistics.~\cite{PhysRevLett.99.126405} This quantity
describes the probability of finding an atom in a particular
many-body state and the expectation value of any local operator
can be easily obtained from it. 
We display schematically the atomic states important for the
forthcoming discussion in Fig.~\ref{fig_splitting}.

\begin{figure}[ht]
\includegraphics[width=\columnwidth]{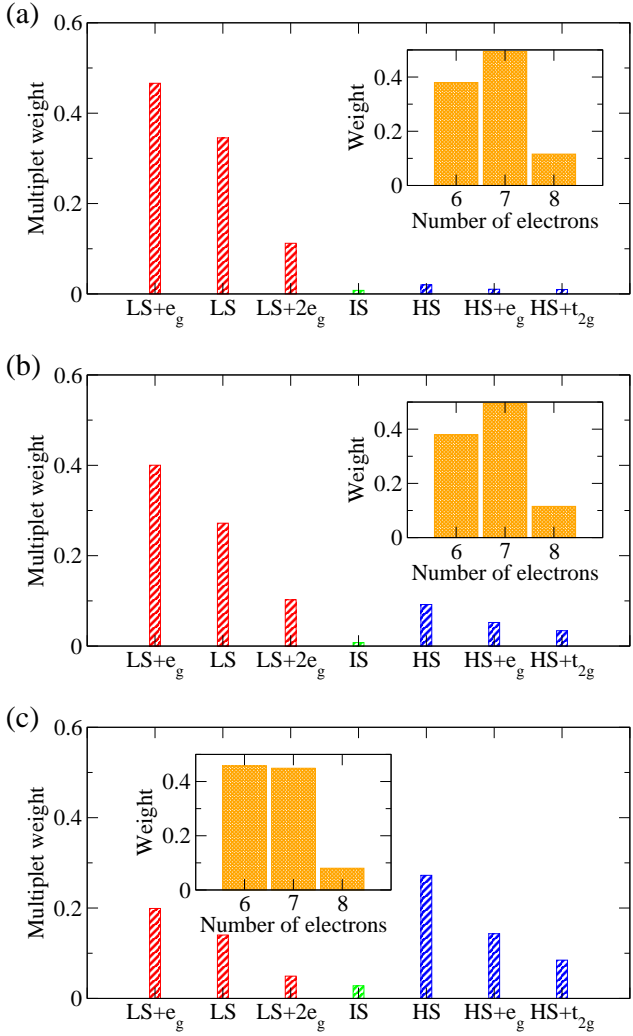}
\caption{
(Color online)
Weight of dominant atomic states for
$U=6.0$~eV, $J=0.8$~eV.
(a) Nonmagnetic solution [A in Fig.~\ref{fig_chi_mtm}(b), $T=580$~K, $\tau_\mathrm{lattice}=5$~K],
(b) low-$T$ solution [B in Fig.~\ref{fig_chi_mtm}(b), $T=580$~K, $\tau_\mathrm{lattice}=450$~K],
(c) high-$T$ solution [C in Fig.~\ref{fig_chi_mtm}(b), $T=1160$~K, $\tau_\mathrm{lattice}=750$~K].
}
\label{fig_stat2}
\end{figure}

There are many atomic states with non-negligible weights
contributing to the partition function
(Fig.~\ref{fig_stat2}).
The total contributions of different charge states (insets
of Fig.~\ref{fig_stat2}) point to sizable valence fluctuations,
which is related to finding substantial admixtures
of $\mathrm{d^7\underline{L}}$ and $\mathrm{d^8\underline{L}^2}$ state
to the $\mathrm{d^6}$ ground state in the cluster calculations.\cite{saitoh}
Unlike the cluster calculations, in DMFT the ligand hole does not
remain coherent with the central Co atom due to the influence of the
rest of the crystal. Therefore, one cannot use the CoO$_6$ eigenstates
to analyze the local dynamics. Instead, we use the statistical
description and also analyze the temporal evolution of the atomic states.

Clearly, the $\mathrm{d^6}$ atomic multiplets denoted as LS, IS, and HS in Fig.~\ref{fig_stat2}
are not sufficient to describe the local physics in \laco. 
We distribute the atomic states into LS, IS, and HS
blocks (Fig.~\ref{fig_splitting}). Although an {\it a priori} assignment
of the blocks is not unique we later present an {\it a posteriori}
justification of our choice. For example, the state denoted as HS+$t_{2g}$ can,
in principle, be reached by adding a $t_{2g}$ electron to the HS state as well as by adding
an $e_g$ electron to the IS state. 
In Fig.~\ref{fig_stat2} we present the state statistics for various lattice parameters and 
temperatures corresponding to the solutions denoted as A, B, and C in Fig.~\ref{fig_chi_mtm}(b).
Besides substantial weights of the $\mathrm{d^7}$ states the figure reveals that
the increasing local moment response is connected to the growing weight of the HS block,
while the IS block has only minor weight.
\begin{figure}[t!]
\includegraphics[width=\columnwidth]{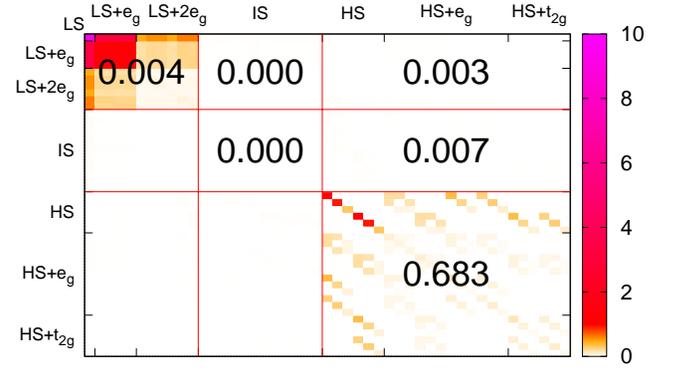}
\caption{
(Color online)
Correlation matrix $\Pi_{\mu\nu}$ between the dominant atomic states
for $T=580$~K and $\tau_\mathrm{lattice}=450$~K [solution B of Fig.~\ref{fig_chi_mtm}(b)].
Color-coded values show the state-by-state relative
contributions in \% to the sum over all pairs. The
numbers within blocks of atomic states indicate the contribution of the blocks
to the local susceptibility in the units of $\mathrm{10^{-3}emu\cdot Oe^{-1}\cdot mol^{-1}}$.
}
\label{fig_chi_matrix1}
\end{figure}

\begin{figure}[t!]
\includegraphics[width=\columnwidth,clip]{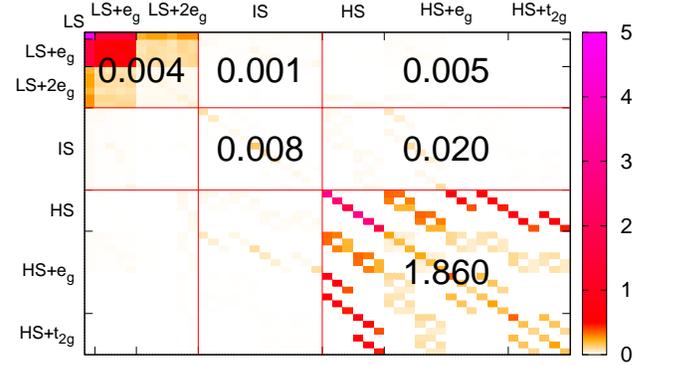}
\caption{
(Color online)
Same as Fig.~\ref{fig_chi_matrix1} for 
$T=1160$~K and $\tau_\mathrm{lattice}=750$~K [solution C of Fig.~\ref{fig_chi_mtm}(b)].
}
\label{fig_chi_matrix2}
\end{figure}

{\it Correlation matrix of local states.}
Although we have identified the atomic states with large weights, 
a question arises whether their appearance is due to a unitary evolution
or rather due to statistical averaging. This question
is closely connected to the decomposition of susceptibility into atomic
states contributions [Eqs.~(\ref{CPi}) and (\ref{eq_chi})]. States connected by a unitary evolution
lead to a large off-diagonal element of the time averaged state-state 
correlation matrix $\Pi$ and thus their individual
contributions to the susceptibility cannot be well defined. On the other hand,
if the weights of different states originate in statistical averaging the
corresponding off-diagonal element of $\Pi$ is vanishing as is its contribution
to the susceptibility.
Analysis of the correlation matrix $\Pi_{\mu\nu}$,
displayed in Figs.~\ref{fig_chi_matrix1} and~\ref{fig_chi_matrix2}, reveals
nonzero off-diagonal elements, indicating a unitary evolution
between the corresponding states. 
Nevertheless, the matrices can be arranged in a block diagonal form, which
justifies our choice of the LS, IS, and HS blocks.
As the unitary evolution between the most populated LS and HS blocks 
has a vanishing probability,
their simultaneous population is a result of the statistical averaging
and blocks generalize the notion of atomic multiplets in isolated atoms.

\begin{figure}
\includegraphics[height=0.5\columnwidth,angle=270,clip]{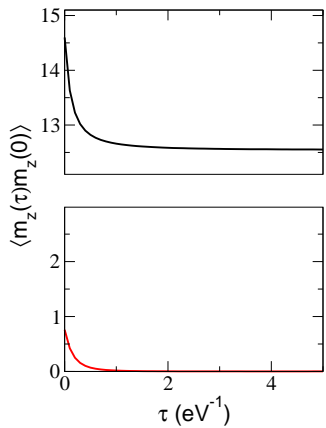}
\caption{(Color online)
HS (upper panel) and LS (lower panel) normalized block
contributions to the spin-spin correlation function,
$\sum_{\mu,\nu\in \mathrm{block}}m_z(\mu )m_z(\nu )C_{\mu\nu}(\tau)/\sum_{\mu,\nu \in \mathrm{block}}\Pi_{\mu\nu}$,
for $T=1160$~K and $\tau_\mathrm{lattice}=750$~K [solution C of Fig.~\ref{fig_chi_mtm}(b)].
}
\label{fig_sts}
\end{figure}

The block-summed contributions to the local susceptibility are indicated
by numbers inside the respective blocks in Figs.~\ref{fig_chi_matrix1} and
\ref{fig_chi_matrix2}.
In both cases the contribution of the HS diagonal block amounts around 97\%
of the total susceptibility. Inspecting the 
block contributions to the spin-spin correlation (Fig.~\ref{fig_sts})
we find a finite $\tau$-constant part of the HS contribution, 
leading to Curie-type susceptibility, in contrast to the rapidly
decaying LS contribution. This allows us to define an effective
HS moment as
\begin{equation}
\mu_{\text{HS}}=\frac{
\sqrt{
\sum_{\mu\nu\in \mathrm{HS}}
m_z(\mu)m_z(\nu)\Pi_{\mu\nu}
}
}
{
\sum_{\mu\nu\in \mathrm{HS}}\Pi_{\mu\nu}
}.
\end{equation}
We obtain $\mu_{\text{HS}}$ of 3.52 and 3.56~$\mu_B$ in the low-$T$ and 
the hight-$T$ solutions, respectively. 
The weak $T$ dependence of the effective moment and its dominant contribution
to the susceptibility $\chi$ justifies expressing $\chi$ as a product
of Curie term $\mu^2/T$ and a $T$-dependent weight. Covalent
Co-O bonding results in about 10\% reduction of the effective
moment from its atomic value of 4~$\mu_B$.

\subsection{Role of $U$, $J$, double counting correction}
Since the form and construction of the Hamiltonian (1) is to some
extent an {\it ad hoc} procedure it is important to 
understand the sensitivity of our conclusions to the particular values
of $U$, $J$, and the double counting energy [$E_\mathrm{dc}$ is the second
term on right-hand side~of Eq.~(\ref{eq_edc})]. Although these are not adjustable
parameters in the present theory, their determination is not unique, which
holds in particular for the double counting correction $E_\mathrm{dc}$.
\begin{figure}[ht]
\includegraphics[width=\columnwidth,clip]{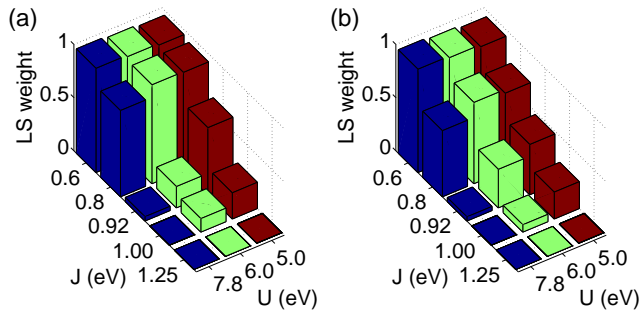} 
\caption{
(Color online)
Weight of the LS block of atomic states (LS, $\mathrm{LS}+e_g$, and $\mathrm{LS}+2e_g$)
for various 
$U$ and $J$ at $\tau_\mathrm{lattice}=5$~K, (a) $T=580$~K, (b) $T=1160$~K.
}
\label{fig_UJ}
\end{figure}
In Fig.~\ref{fig_UJ} we show the dependence of the weight of the LS block
for various values of $U$ and $J$. As expected $J$ is the more important
parameter. Its cDFT value falls into the spin state crossover range of
0.8\,--\,0.9~eV. Outside this range the results are insensitive to temperature
or the variation of $U$. Variation of $U$ inside the crossover regime
has some impact as higher $U$ suppresses the fluctuations  
to the $\mathrm{d^7\underline{L}}$ and $\mathrm{d^8\underline{L}^2}$ states.
As a result the LS state is destabilized.
\begin{figure}[ht]
\includegraphics[angle=270,width=\columnwidth,clip]{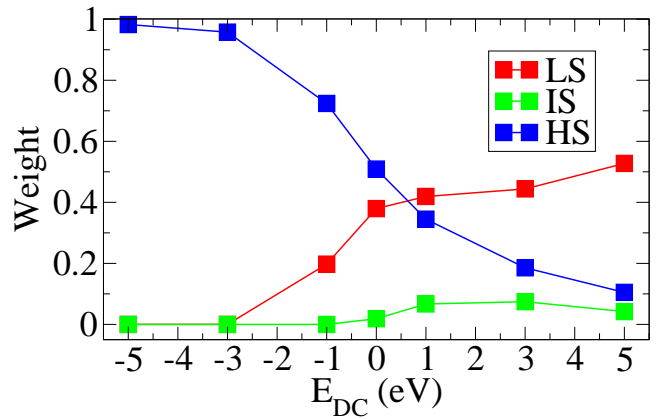}
\caption{
(Color online)
Weight of LS, IS, and HS blocks of atomic multiplets for various
double counting energies for high-$T$ solution
($U=6$~eV, $J=0.8$~eV, $T=1160$~K, and $\tau_\mathrm{lattice}=750$~K).
The values indicate the offset from the self-consistent value of 33.56~eV.
}
\label{fig_eDC}
\end{figure}

We have examined the role of double counting correction ($E_\mathrm{dc}$) for
$U=6.0$~eV, $J=0.8$~eV, $T=1160$~K, and $\tau_\mathrm{lattice}=750$~K.
We varied $E_\mathrm{dc}$ by $\delta E_\mathrm{dc}$ in the range of $\pm 5$~eV
around the self-consistent value of 33.56~eV.
The positive values of $\delta E_\mathrm{dc}$ mean that the Co:$d$ states
are shifted downward in energy closer to the O:$p$
state, which in turn enhances the hybridization and leads to
preference of a LS metallic phase.
The weights of the LS, IS and HS blocks 
are shown in Fig.~\ref{fig_eDC}. For $\delta E_\mathrm{dc}>1$
the system becomes metallic
and the definition of the LS, IS, and HS blocks loses
its justification. 

Based on the one-particle spectra and the overall behavior
of our results we conclude that the self-consistent value of $E_\mathrm{dc}$
provides a rather good description of the actual material.
We also point out the $T$-dependent variation of $E_\mathrm{dc}$
is rather small and had minor effect on the $T$ dependence of
both $\chi$ and $\Delta_{\text{eff}}$.

\section{Conclusion}
We have studied the temperature dependence of the magnetic
and spectral properties of \laco\ using the LDA+DMFT approach. Our results show that
the local moment response at elevated temperatures is associated
with the HS state of Co ion and that there is an effective interatomic
repulsion between HS atoms in the LS matrix. Our findings at this point agree
with the VCA calculations \cite{eder} and LDA+U cluster calculations \cite{knizek1}
as well as with the conclusions of the experimental studies.~\cite{kyomen1,kyomen2,haverkort}
On the other hand, our results are inconsistent with interpretation
of the intermediate temperature phase in terms of the IS state.~\cite{korotin,radaelli,zobel}
Furthermore, since a purely electronic mechanism
of the HS-LS attraction exists \cite{krapek} the absence or smallness 
of the breathing mode distortion \cite{radaelli}
does not exclude HS-LS short range ordering.
The experimentally observed (anomalous) lattice expansion has
a pronounced effect on the LS-HS crossover, which leads to the conclusion
that the lattice provides a strong positive feedback for the 
crossover.

To account for the strong covalent bonding 
with O the notion of Co HS state has to be generalized 
to include not only $d^6$, but also $d^7$ and $d^8$ electron configurations.
This leads
to reduction of the magnitude of the effective HS moment. Therefore,
using the apparent magnitude of the magnetic moment as an indicator
of the underlying atomic state may lead to incorrect conclusions.

Although we have varied the computational parameters in a wide range
we have not found a phase that would be dominated by the IS state.
Therefore, scenarios invoking HS-IS crossover \cite{knizek1,kyomen2} 
are not consistent with our results. Guided by a similar LDA+DMFT study
on metallic SrCoO$_3$, it is plausible that in the  metallic
phase of \laco\ observed experimentally above 600~K distinction between IS and HS state
is not possible as those are connected by unitary evolution of the system.

Based on the above observations we propose the following scenario of \laco\
physics, which is in many aspects similar to Refs.~\onlinecite{knizek1,eder}.
At the lowest temperatures most Co ions are in the LS state with isolated Co ions in the 
excited HS state. Increasing temperature assisted by the lattice feedback
leads to growing density of the HS sites. Effective repulsion
keeps the HS sites apart leading to a short range HS-LS order, which is 
responsible for the insulating behavior in the 100--500~K range, similar
to the observation made in Ref.~\onlinecite{krapek}. We speculate that the second
crossover experimentally observed around 500~K is associated with ``melting" of the LS-HS order.
This leads to an anomalous lattice expansion due to the breaking of attractive
LS-HS bonds. The experimentally observed onset of metallicity 
changes the local moment character by coherently admixing some
IS-like states to the dominant HS configuration. The distinction
between HS and IS in the high $T$ metallic phase is thus not possible.

\subsection*{Acknowledgments}
We thank Z. Jir\'ak for numerous discussion and suggestions
concerning the manuscript.
This work was supported by Grant No.~P204/10/0284 of the
Grant Agency of the Czech Republic, by the Deutsche
Forschungsgemeinschaft through FOR1346,
by the Russian Foundation for Basic Research (Project Nos.~10-02-00046 and 10-02-96011) and by the Fund of the President of the Russian Federation for the Support of Scientific Schools NSH-6172.2012.2.

\appendix*
\section{Comparison of full and density-density interactions}

\begin{figure}[ht]
\includegraphics[angle=270,width=\columnwidth,clip]{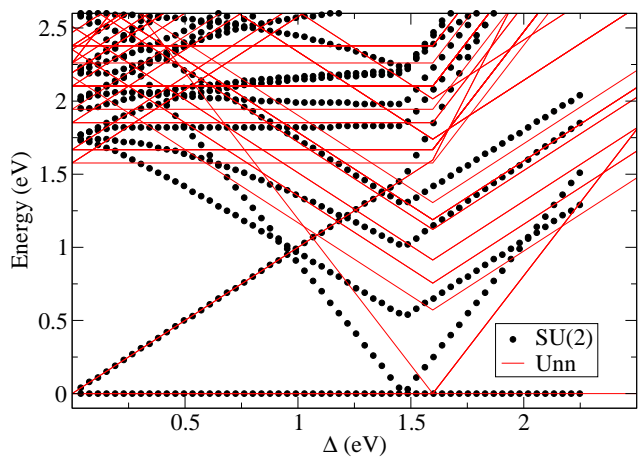}
\caption{
(Color online)
Comparison of Tanabe-Sugano diagrams for the full
rotationally invariant [SU(2); black points] and density-density (Unn; red lines)
Coulomb interaction. Only the low-energy multiplets are displayed for the
crystal field $\Delta$ around the LS and HS multiplet crossing.
}
\label{fig_Tanabe_Sugano}
\end{figure}

The restriction to the density-density terms in the Coulomb interaction
is an approximation which
greatly reduces the computational effort. 
To assess the approximation to the excitation energies
we compare the Tanabe-Sugano~\cite{tanabe-sugano} diagrams
for the $\mathrm{d^6}$ configuration
obtained with the density-density and full Coulomb interaction
in Fig.~\ref{fig_Tanabe_Sugano}.
The largest difference of the two spectra is found at zero
crystal-field splitting where the off-diagonal elements, neglected
in density-density approximation, play the most important
role. At larger crystal fields some of the full multiplet states
(e.g., LS or HS with the maximal spin projection) become dominated by a
single Slater determinant
and their energies are close to the density-density ones.
Importantly, the degeneracies of the multiplets for the two interactions
differ.
Recently, the effect of the density-density interaction was studied
on a similar material SrCoO$_3$ and good agreement was
found for the multiplet-averaged state weights.~\cite{PhysRevLett.109.117206}


\end{document}